\documentclass[sigconf]{acmart}
\copyrightyear{2026}
\acmYear{2026}
\setcopyright{cc}
\setcctype{by}
\acmConference[KDD '26]{Proceedings of the 32nd ACM SIGKDD Conference on Knowledge Discovery and Data Mining V.2}{August 09--13, 2026}{Jeju Island, Republic of Korea}
\acmBooktitle{Proceedings of the 32nd ACM SIGKDD Conference on Knowledge Discovery and Data Mining V.2 (KDD '26), August 09--13, 2026, Jeju Island, Republic of Korea}
\acmDOI{10.1145/3770855.3818099}
\acmISBN{979-8-4007-2259-2/2026/08}

\usepackage{multirow}
\usepackage{booktabs}
\usepackage{enumitem}
\usepackage{balance}
\usepackage{tikz}

\AtBeginDocument{%
  }

\begin{document}

\title{VeriHGN: Heterogeneous Graph–Based Congestion Prediction for Chip Layout Verification}
\author{Runbang Hu}
\affiliation{
  \institution{The University of Texas at Arlington}
  \city{Arlington}
  \state{Texas}
  \country{USA}
}
\email{rxh0841@mavs.uta.edu}

\author{Bo Fang}
\affiliation{
  \institution{The University of Texas at Arlington}
  \city{Arlington}
  \state{Texas}
  \country{USA}
}
\email{bo.fang@uta.edu}
\authornote{Co-corresponding authors}
\author{Bingzhe Li}
\affiliation{
  \institution{The University of Texas at Dallas}
  \city{Dallas}
  \state{Texas}
  \country{USA}
}
\email{bingzhe.li@utdallas.edu}

\author{Yuede Ji*}
\affiliation{
  \institution{The University of Texas at Arlington}
  \city{Arlington}
  \state{Texas}
  \country{USA}
}
\email{yuede.ji@uta.edu}

    \renewcommand{\shortauthors}{Runbang Hu et al.}
    
\newcommand{\sigmoid}{\sigma}
\newcommand{\vect}[1]{\boldsymbol{#1}}
\newcommand{\mat}[1]{\boldsymbol{#1}}
\newcommand*\circled[2][black]{%
  \tikz[baseline=(char.base)]{\node[
    shape=circle, draw=none, thick,
    fill=#1, inner sep=0.9pt] (char)
  {\textcolor{white}{#2}};}}
\begin{abstract}
As Very Large Scale Integration (VLSI) designs continue to scale in size and complexity, layout verification has become a central challenge in modern Electronic Design Automation (EDA) workflows. In practice, congestion can only be accurately identified after detailed routing, making traditional verification both time-consuming and costly. Learning-based approaches have therefore been explored to enable early-stage congestion prediction and reduce routing iterations. However, although prior methods incorporate both netlist connectivity and layout features, they often model the two in a loosely coupled manner and primarily produce numerical congestion estimates. We propose VeriHGN, a verification framework built on an enhanced heterogeneous graph that unifies circuit components and spatial grids into a single relational representation, enabling more faithful modeling of the interaction between logical intent and physical realization. Experiments on industrial benchmarks, including ISPD2015, CircuitNet-N14, and CircuitNet-N28, demonstrate that VeriHGN achieves the best or near-best performance over state-of-the-art methods in prediction accuracy and correlation metrics.
\end{abstract}

\maketitle
\newcommand\kddavailabilityurl{https://doi.org/10.5281/zenodo.20496640}
\ifdefempty{\kddavailabilityurl}{}{
\begingroup\small\noindent\raggedright\textbf{Resource Availability:}\\
The source code and artifacts of this paper have been made publicly available at
\url{\kddavailabilityurl}. The source code repository is available at
\url{https://github.com/SC-Lab-Go/VeriHGN}.

\endgroup
}
\begin{figure}[h]
    \centering
    \includegraphics[width=\linewidth]{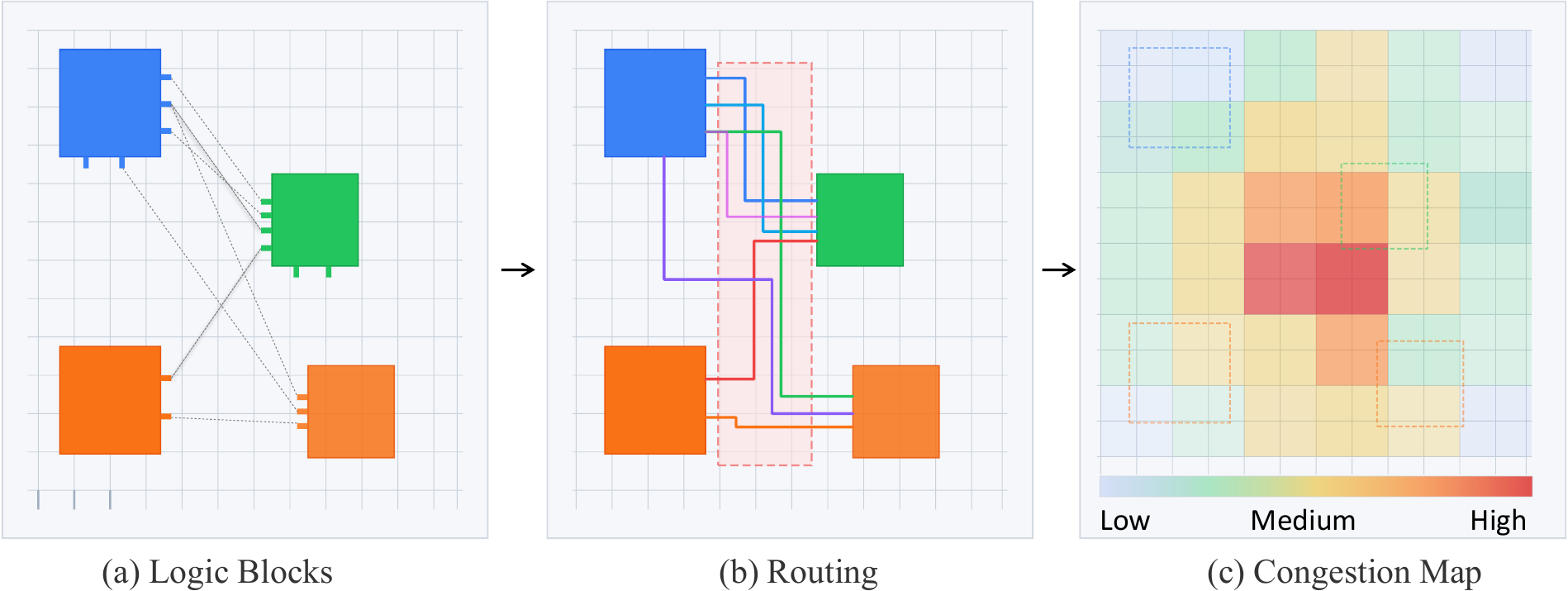}
     \caption{Example of routing congestion formation in a placement–routing workflow. (a): multiple logic blocks are placed on the layout grid with interconnecting nets. (b): global routing paths overlap and compete for limited routing resources, creating localized routing congestion. (c): accumulated routing demand is visualized as a congestion map, where colors indicate congestion levels (low to high), and red regions highlight areas where routing demand exceeds available capacity.}
    \label{fig:congestion_explanation}
\end{figure}
\section{Introduction}
As Very Large Scale Integration (VLSI) designs continue to scale in size and complexity~\cite{mohammad2023ai,alioto2012ultra,saraswat2005effect,chen2020pros}, layout verification has become a critical bottleneck in modern Electronic Design Automation (EDA) workflows~\cite{kuon2005design,shostak1983verification,zhang2006speeding,lin2019dreamplace,pandey2018machine}. Among verification tasks, routing congestion directly affects timing closure, power integrity, and design convergence. In industrial flows, accurate congestion information is typically available only after detailed or global routing, which is computationally expensive and often requires multiple iterations before a design becomes routable. As a result, late discovery of congestion hotspots prolongs turnaround time and increases design cost. Figure~\ref{fig:congestion_explanation} illustrates how congestion arises during routing. For designs with millions of cells, a single physical design iteration can take days to weeks, and convergence often requires repeated iterations. Modern SoC (System-on-chip) designs have scaled to tens of millions of cells~\cite{jiang2023accelerating}. In large-scale physical design, the overall process can take hours to days, with routing being one of the most critical steps and consuming up to 70\% of total design time~\cite{yu2019painting, goswami2021congestion}.

To mitigate this issue, learning-based congestion prediction has emerged as an attractive alternative for early-stage layout verification without costly routing engines~\cite{spindler2007fast,xie2018routenet,kirby2019congestionnet,ghose2021generalizable,wang2022lhnn,yang2022versatile,zhangmihc,cai2025st,zou2025lay,sanchez2023comprehensive}. By predicting congestion from high-level synthesis (HLS) and placement-level information, these methods provide fast feedback to guide downstream optimization and reduce routing iterations. However, despite substantial progress, achieving both high prediction accuracy and practical usefulness for layout improvement remains challenging, especially for large-scale industrial designs.

\textbf{Challenges.}
Despite substantial progress, existing learning-based congestion predictors still struggle to jointly model netlist connectivity and physical layout geometry. Geometry-centric methods are effective at capturing local spatial patterns but have limited ability to reason about long-range routing demand induced by netlist structure. In contrast, netlist-centric methods better preserve circuit connectivity but often lack fine-grained spatial context required for congestion estimation in placed designs. Even multi-view approaches typically process layout and topology through separate encoders with late-stage fusion, which limits fine-grained interaction between logical connectivity and physical realization.

This limitation is critical because routing congestion is not merely a regression problem over spatial features or netlist statistics. Rather, it emerges from the interaction among logical connectivity, physical placement, and multi-scale routing resource competition. Local geometric density alone cannot explain hotspots caused by long-span or high-fanout nets, while purely topological representations overlook constraints from routing grids and local placement competition. Therefore, effective congestion prediction requires a unified representation that enables information exchange between circuit structure and spatial layout across multiple granularities.

To address these challenges, we propose VeriHGN, a unified heterogeneous graph-based framework for early-stage routing congestion prediction. VeriHGN represents cells, nets, and spatial regions as distinct node types within a single relational graph, enabling direct interaction between netlist topology and layout geometry. Spatial regions are further organized into a hierarchical grid, allowing congestion information to propagate across neighboring regions and multiple resolutions. By coupling netlist connectivity with multi-scale spatial context, VeriHGN captures both local congestion from dense placement and regional congestion induced by long-span or high-fanout nets.

% \textbf{Contribution.}
% To address these challenges, we propose VeriHGN, a congestion prediction framework for chip layout verification based on an enhanced heterogeneous graph representation. Unlike prior approaches that treat netlist and layout as loosely connected modalities, VeriHGN unifies circuit components and spatial grids into a single relational graph, enabling direct modeling of interactions between logical intent and physical realization. This unified formulation allows congestion to be analyzed as a structural property of the joint netlist–layout system rather than a purely spatial or topological artifact.
% Extensive experiments on public industrial benchmarks, including ISPD2015, CircuitNet-N14, and CircuitNet-N28, demonstrate that VeriHGN consistently outperforms state-of-the-art methods in prediction accuracy and correlation metrics, while simultaneously providing interpretable insights that support efficient, early-stage layout verification.
% \textbf{Contributions.}
This paper makes three main contributions.
\begin{itemize}[leftmargin=*, topsep=2pt]
  \item We propose \textbf{VeriHGN}, a heterogeneous graph-based framework that unifies circuit netlist and spatial routing grids into a single representation for early-stage chip layout verification.

  \item We design a multi-resolution heterogeneous message passing architecture that explicitly models interactions among cells, nets, and hierarchical grids, enabling joint reasoning over logical connectivity and physical layout constraints.

  \item Experiments on ISPD2015~\cite{bustany2015ispd}, CircuitNet-N14, and CircuitNet-N28~\cite{2024circuitnet}  show that VeriHGN achieves state-of-the-art hotspot-oriented congestion prediction, with the best cell-level rank correlations across all benchmarks and best or near-best grid-level performance in most settings.

\end{itemize}

\section{Background and Related Work}

\subsection{Learning-Based Congestion Prediction}

Learning-based congestion prediction estimates routing congestion from pre-routing design representations to reduce expensive routing iterations~\cite{spindler2007fast,chen2020pros,pandey2018machine,sanchez2023comprehensive}. Existing methods can be broadly categorized into geometry-based and topology-based approaches. Geometry-based methods rasterize layouts into grid-level feature maps and apply convolutional or transformer-based models~\cite{xie2018routenet,cai2025st,zou2025lay}. For example, RouteNet~\cite{xie2018routenet} encodes density and RUDY features as multi-channel images, while ST-FPN~\cite{cai2025st} and Lay-Net~\cite{zou2025lay} improve spatial modeling with Swin Transformer or vision-transformer backbones. These methods benefit from efficient dense tensor operations and strong local spatial modeling, but rasterization may obscure netlist topology and long-range routing demand~\cite{kirby2019congestionnet,ghose2021generalizable,wang2022lhnn,yang2022versatile}.

Topology-based methods represent circuits as graphs or hypergraphs and use GNNs to capture connectivity-driven routing pressure~\cite{kipf2016semi,velivckovic2017graph,xu2018powerful,sanchez2023comprehensive}. CongestionNet~\cite{kirby2019congestionnet}, CircuitGNN~\cite{yang2022versatile}, and MIHC~\cite{zhangmihc} respectively model cell-level connectivity, bipartite cell-net graphs, and multi-view topology-geometry information. While these methods better preserve circuit structure, many multi-view designs still rely on separate encoders and late fusion, limiting fine-grained interaction between physical layout and logical connectivity~\cite{ghose2021generalizable,wang2022lhnn,zhangmihc,zou2025lay}. In contrast, VeriHGN unifies cells, nets, and hierarchical grid regions in a single heterogeneous graph.

\subsection{Heterogeneous Graph Neural Networks}

GNNs have been widely used for graph representation learning, including node classification, graph classification, and link prediction~\cite{kipf2016semi,hamilton2017inductive,zhang2018link,dwivedi2020benchmarking}. Heterogeneous GNNs extend standard message passing to graphs with multiple node and edge types~\cite{schlichtkrull2018modeling,wang2019heterogeneous,hu2020heterogeneous,yang2023simple}. Unlike homogeneous GNNs that use shared aggregation rules~\cite{kipf2016semi,velivckovic2017graph,xu2018powerful}, heterogeneous models learn type- or relation-specific transformations to preserve semantic differences. Representative models such as R-GCN~\cite{schlichtkrull2018modeling}, HGAN~\cite{wang2019heterogeneous}, HGT~\cite{hu2020heterogeneous}, and Simple-HGN~\cite{yang2023simple} capture heterogeneous interactions through relation-specific weights, hierarchical attention, or type-aware attention. This capability is particularly relevant to EDA graphs, where cells, nets, pins, and spatial regions correspond to distinct physical entities~\cite{sanchez2023comprehensive,yang2022versatile,wang2022lhnn,zhangmihc}.

However, directly applying general heterogeneous GNNs to chip layout verification remains challenging. Circuit graphs are large, degree distributions are highly skewed due to high-fanout nets and macros, and edge types encode different routing semantics, including logical connectivity, local placement competition, and spatial resource aggregation~\cite{ghose2021generalizable,wang2022lhnn,yang2022versatile,zhangmihc}. VeriHGN addresses these challenges with lightweight relation-specific message passing tailored to congestion prediction, explicitly modeling cell-net connectivity, local geometric interaction, grid-net coupling, and hierarchical grid propagation.

\begin{figure*}[t]
    \centering
    \includegraphics[width=0.96\textwidth]{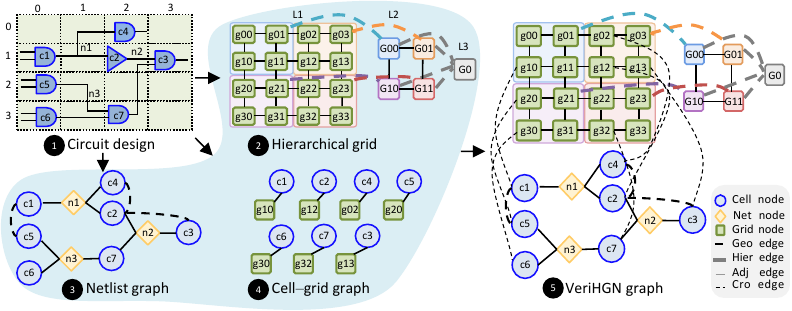}
    \caption{Overview of VeriHGN graph construction. Starting from the original circuit design (\circled{1}), we derive a hierarchical grid representation (\circled{2}) for multi-level spatial structure, a netlist graph (\circled{3}) for circuit connectivity, and a cell-to-grid relationship graph (\circled{4}) for spatial assignment. These components are integrated into a unified heterogeneous graph (\circled{5}), enabling joint structural and spatial message passing.}
    \label{fig:graph_construction}
\end{figure*}

\subsection{Multi-Scale Spatial Modeling for Routing Congestion}

Routing congestion is inherently multi-scale~\cite{spindler2007fast,xie2018routenet,wang2022lhnn,cai2025st,zou2025lay}. Local hotspots are often caused by dense placement, pin access contention, and limited routing resources within small neighborhoods, while regional congestion can arise from long-span nets and accumulated routing demand across multiple regions. Therefore, effective congestion prediction requires both fine-grained spatial sensitivity and broader regional context.

Most geometry-based methods represent layouts as fixed-resolution grids and apply convolutional or transformer-based encoders~\cite{xie2018routenet,cai2025st,zou2025lay}. Although effective for local density modeling, a single resolution may lose detail when too coarse or become expensive when too fine. Some methods introduce multi-scale context through image pyramids, feature pyramid networks, or hierarchical aggregation~\cite{ronneberger2015u,cai2025st,zou2025lay}, but the spatial hierarchy is usually implicit in the neural architecture. VeriHGN instead treats multi-resolution grid regions as first-class nodes in the heterogeneous graph, enabling explicit message passing across spatial scales and direct interaction between layout hierarchy and netlist connectivity.

\section{Methodology}

\subsection{Congestion Prediction Graph Construction}

We construct a heterogeneous congestion prediction graph that explicitly encodes both circuit connectivity and spatial routing demand. Figure~\ref{fig:graph_construction} illustrates the overall construction process. The key design principle is to expose the primary sources of routing congestion, including logical fanout, geometric proximity, and regional resource contention, as first-class graph entities. Instead of treating netlist and layout as separate modalities, our formulation integrates them into a single relational graph, allowing congestion to be modeled as an emergent property of their interaction.

\textbf{Node types.}
The graph contains three node types. \emph{Cell nodes} correspond to standard cells and macros, representing instance-level routing demand and serving as prediction targets for cell congestion. \emph{Net nodes} correspond to electrical nets and aggregate routing demand induced by their incident cells. \emph{Grid nodes} correspond to spatial regions obtained by discretizing the placement area and are organized hierarchically to represent regional routing resources. This tripartite design captures the multi-origin nature of congestion: local demand from cell geometry and pin access, global routing pressure from multi-terminal and long-span nets, and regional imbalance in routing resources.

\textbf{Hierarchical grid construction.}
To capture spatial routing patterns at different scales, we construct grid nodes at multiple resolutions. At the finest level (L0), the placement area is divided into an $M \times N$ fine grid, where each grid cell corresponds to one grid node and captures local routing density and resource contention. Coarser levels are built by spatial aggregation: each level-$\ell{+}1$ grid node represents a non-overlapping $2 \times 2$ block of level-$\ell$ grid nodes. All grid levels are included in the graph, enabling the model to capture both localized hotspots and regional congestion caused by routing demand spanning multiple tiles.

\textbf{Edge types.}
Edges encode logical and spatial relationships. \emph{Pin edges} connect cells and nets according to netlist incidences. \emph{Geometric cell edges} connect spatially nearby cells to model local competition for routing resources. \emph{Grid adjacency edges} connect grid nodes within the same resolution using 4-connected neighborhoods. \emph{Hierarchical grid edges} connect each coarse grid node to its corresponding fine grid nodes, encoding parent--child relationships across resolutions. Together, these edges form a unified graph that jointly represents instance-level connectivity, local spatial interaction, and multi-scale regional routing demand.

% Figure~\ref{fig:overview} illustrates the overall pipeline. Starting from raw placement and netlist data, we first construct enriched node and edge attributes (Section~\ref{sec:features}). We then perform pin-aware and geometry-aware message passing on the unified heterogeneous graph (Section~\ref{sec:pin_aware}). To efficiently capture large-scale spatial context, spatial-bin nodes are further refined by a multi-resolution hierarchy (Section~\ref{sec:hierarchy}) before interacting with other node types. Finally, a congestion-aware training objective is applied to guide optimization (Section~\ref{sec:training}).

\subsection{Enriched Feature Engineering}
\label{sec:features}

The heterogeneous graph is parameterized by node and edge level attributes that encode geometric, topological, and semantic information available from the placed design. Rather than relying on deep propagation to implicitly recover routing demand signals, we explicitly construct informative features for different graph components. These features provide strong inductive biases and improve sample efficiency on limited benchmark designs.

\textbf{Cell node feature.} Each cell $v$ is represented by a 13-dim vector:
\begin{equation}
{x}_v^{\text{cell}} =
\big[\underbrace{x, y, w, h, a}_{\text{geometry}},\;
\underbrace{d_{\text{net}}, p, \bar{l}, \hat{l}, \bar{d}}_{\text{topology}},\;
\underbrace{m, r_{\text{in}}, f_t}_{\text{type}}\big],
\label{eq:cell_feat_rewrite}
\end{equation}
where $(x,y)$ are placement coordinates, $(w,h)$ are cell dimensions, and $a=w\cdot h$ is area.
Topological statistics include the incident-net count $d_{\text{net}}$, pin count $p$, the mean and max half-perimeter wire length (HPWL) of incident nets $(\bar{l}, \hat{l})$, and the mean incident-net degree $\bar{d}$.
Type-related attributes include a macro indicator $m$, input-pin ratio $r_{\text{in}}$, and normalized type frequency $f_t$ (fraction of cells sharing the same master). Continuous features are standardized.

\textbf{Net node features.} Each net $e$ is initialized by a 6-dim vector:
\begin{equation}
{x}_e^{\text{net}} =
\big[\text{HPWL}(e),\; \Delta x(e),\; \Delta y(e),\; |e|,\; \log|e|,\; A_{\text{bbox}}(e)\big],
\label{eq:net_feat_rewrite}
\end{equation}
where $\Delta x,\Delta y$ denote bounding-box spans, $|e|$ is net degree, and $A_{\text{bbox}}=\Delta x\cdot\Delta y$ is bounding-box area. The logarithmic term stabilizes heavy-tailed degree distributions.

\textbf{Grid node features.}
Each grid node $g$ corresponds to a spatial bin at a given resolution level and aggregates local placement statistics within its spatial extent. We initialize each grid node with a feature vector:
\begin{equation}
\vect{x}_g^{\text{grid}} =
\big[\rho_{\text{cell}}(g),\; \rho_{\text{pin}}(g),\; \rho_{\text{net}}(g),\;
\bar{a}_{\text{cell}}(g),\; \bar{\ell}_{\text{net}}(g)\big],
\label{eq:grid_feat}
\end{equation}
where $\rho_{\text{cell}}(g)$, $\rho_{\text{pin}}(g)$, and $\rho_{\text{net}}(g)$ denote the normalized densities of cells, pins, and nets intersecting grid $g$, respectively.
$\bar{a}_{\text{cell}}(g)$ is the mean cell area within the grid, and
$\bar{\ell}_{\text{net}}(g)$ is the mean HPWL of nets overlapping the grid region.
The same feature definition is applied consistently across fine and coarse grid levels, enabling multi-resolution spatial reasoning.

\textbf{Edge features.}
Each edge type is associated with its own feature representation. \emph{Pin edges} $(v,e)$ encode cell--net incidences and are initialized with a feature vector:
\begin{equation}
{x}_{v \to e}^{\text{pin}} =
\big[\text{dir}(v,e),\; \delta_x(v,e),\; \delta_y(v,e)\big],
\label{eq:pin_feat}
\end{equation}
where $\text{dir}(v,e)\in\{0,1\}$ indicates signal direction (input or output), and $(\delta_x,\delta_y)$ denote the normalized offset of the cell location relative to the centroid of the net bounding box.

\emph{Geometric cell edges} $(v_i,v_j)$ encode spatial proximity between neighboring cells and are initialized as:
\begin{equation}
{x}_{ij}^{\text{geom}} =
\big[\Delta x_{ij},\; \Delta y_{ij},\; d_{\text{manh}}(v_i,v_j),\; d_{\text{eucl}}(v_i,v_j)\big],
\label{eq:geom_feat}
\end{equation}
where $(\Delta x_{ij},\Delta y_{ij})$ denote relative displacement, and $d_{\text{manh}}$, $d_{\text{eucl}}$ are the Manhattan and Euclidean distances between cells.

\emph{Hierarchical grid edges} connect grid nodes across adjacent resolution levels. These edges are unweighted and do not carry explicit features; instead, they serve as structural links that propagate spatial context between fine and coarse grids during message passing.

\subsection{Relation-Specific Message Passing}
\label{sec:message_passing}

\begin{figure}[t]
    \centering
    \includegraphics[width=\linewidth]{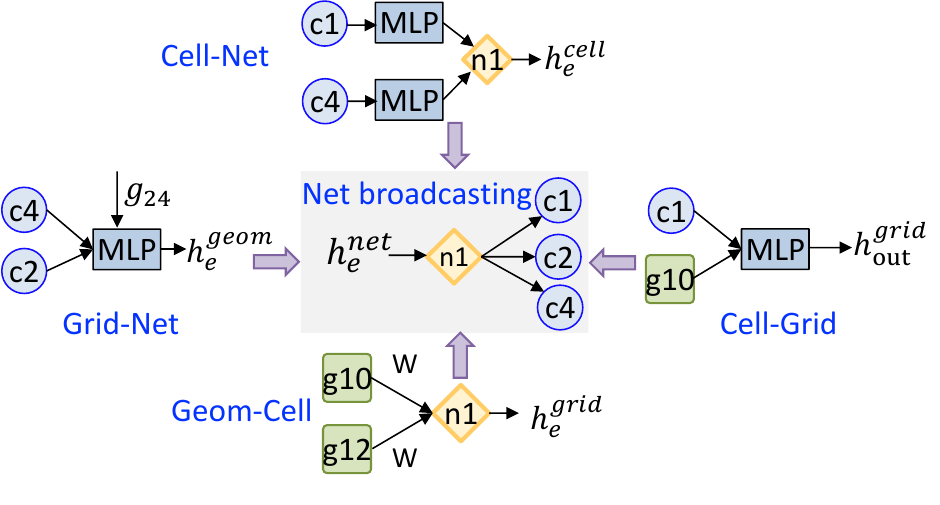}
    \caption{Relation-based message passing. Messages are propagated along different relation types (cell--net, grid--net, geom--cell, and cell--grid) using relation-specific MLPs, enabling information exchange across circuit components and spatial grids.}
    \label{fig:relational_message_passing}
\end{figure}

Figure~\ref{fig:relational_message_passing} illustrates the relation-specific message passing process. Each edge type defines a distinct message function, allowing routing-demand signals to propagate through both circuit connectivity and spatial relations.

\textbf{Cell-net message passing.}
For each pin edge $(v,e)$ connecting a cell node $v$ to a net node $e$, we compute a pin-aware message conditioned on the cell embedding and pin attributes:
\begin{equation}
\vect{m}_{v \to e}
=
\text{MLP}_{\text{pin}}
\big(
[\vect{h}_v \;\|\; \vect{p}_{v \to e}]
\big),
\label{eq:cell_to_net_msg}
\end{equation}
where $\vect{p}_{v \to e}$ encodes signal direction and relative pin offset. Messages from incident cells are averaged to form the cell-induced net representation. Similarly, grid-to-net, net-to-cell, and cell-to-grid relations are modeled using relation-specific linear projections followed by mean aggregation:
\begin{equation}
\vect{h}_i^{(r)}
=
\frac{1}{|\mathcal{N}_r(i)|}
\sum_{j \in \mathcal{N}_r(i)}
\mat{W}_r \vect{h}_j,
\label{eq:relation_agg}
\end{equation}
where $r$ denotes a relation type and $\mathcal{N}_r(i)$ is the corresponding neighbor set. This unified formulation transfers routing demand between cells, nets, and grid tiles while preserving relation-specific semantics.

\textbf{Geometric cell--cell message passing.}
For each geometric edge $(i,j)$ connecting nearby cells, we apply a gated aggregation to modulate neighbor influence based on spatial relevance:
\begin{equation}
\alpha_{ij}
=
\sigmoid
\Big(
\mat{W}_g
[\vect{h}_i \;\|\; \vect{h}_j \;\|\; \vect{g}_{ij}]
\Big),
\label{eq:geom_gate}
\end{equation}
where $\vect{g}_{ij}$ encodes relative displacement and distance features. The geometric message for cell $i$ is then computed as:
\begin{equation}
\vect{h}_i^{\text{geom}}
=
\frac{1}{|\mathcal{N}_g(i)|}
\sum_{j \in \mathcal{N}_g(i)}
\alpha_{ij}\,
\mat{W}_g'\vect{h}_j.
\label{eq:geom_agg}
\end{equation}
This mechanism captures local routing competition while suppressing spatially close but congestion-irrelevant neighbors.

\textbf{Node state updates.}
Cell and grid embeddings are updated by fusing their original embeddings with relation-specific messages through residual transformations. For cell nodes, the update is:
\begin{equation}
\vect{h}_v'
=
\text{ReLU}
\Big(
\mat{W}_f
[\vect{h}_v \;\|\; \vect{h}_v^{\text{net}} \;\|\; \vect{h}_v^{\text{geom}}]
\Big)
+
\vect{h}_v,
\label{eq:cell_update}
\end{equation}
where the residual connection improves training stability and mitigates oversmoothing. Grid nodes are updated analogously by combining their original embeddings with aggregated cell and grid-level messages.

\subsection{Hierarchical Grid Representation Learning}
\label{sec:grid_refine}

\begin{figure}[t]
    \centering
    \includegraphics[width=\linewidth]{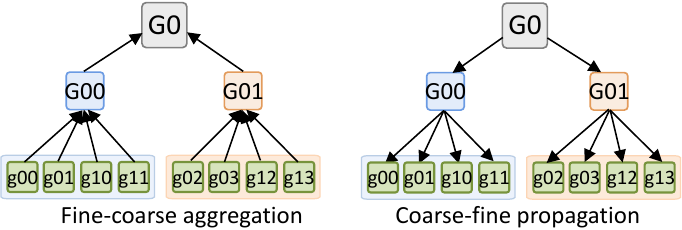}
    \caption{Hierarchical grid representation learning. Grid features are aggregated from fine to coarse levels and then propagated back from coarse to fine levels, enabling multi-scale spatial information exchange.}
    \label{fig:grid_representation_learning}
\end{figure}

Grid node embeddings are learned on the hierarchical grid subgraph to capture spatial routing patterns across multiple resolutions. As shown in Figure~\ref{fig:grid_representation_learning}, grid nodes are organized into fine and coarse levels, where coarse grids summarize larger spatial regions and fine grids preserve local detail. The hierarchy enables efficient propagation of global spatial context while maintaining fine-grained congestion sensitivity.

\textbf{Fine-to-coarse aggregation.}
Let $\mathcal{V}_{\text{grid}}^{(\ell)}$ denote the set of grid nodes at level $\ell$, with $\ell=0$ corresponding to the finest grid.
Each coarse grid node $g^{(\ell+1)}$ aggregates information from its associated fine-grid children
$\mathcal{C}(g^{(\ell+1)}) \subset \mathcal{V}_{\text{grid}}^{(\ell)}$:
\begin{equation}
{h}_{g}^{(\ell+1)}
=
\phi_{\text{fc}}^{(\ell)}
\left(
\frac{1}{|\mathcal{C}(g)|}
\sum_{g' \in \mathcal{C}(g)}
{h}_{g'}^{(\ell)}
\right),
\label{eq:fine_to_coarse}
\end{equation}
where $\vect{h}_{g'}^{(\ell)}$ denotes the grid embedding at level $\ell$, and
$\phi_{\text{fc}}^{(\ell)}(\cdot)$ is a learnable transformation (implemented as a linear layer followed by nonlinearity).
This bottom-up aggregation allows each coarse grid node to summarize routing demand over an increasingly large spatial region.

\textbf{Coarse-to-fine propagation.}
After aggregation at coarse levels, spatial context is propagated back to finer grids through the hierarchical parent--child relations.
For each fine grid node $g^{(\ell)}$, we propagate information from its parent node $\pi(g^{(\ell)}) \in \mathcal{V}_{\text{grid}}^{(\ell+1)}$:
\begin{equation}
\tilde{\vect{h}}_{g}^{(\ell)}
=
\phi_{\text{cf}}^{(\ell)}
\big(
\vect{h}_{\pi(g)}^{(\ell+1)}
\big),
\label{eq:coarse_to_fine}
\end{equation}
where $\phi_{\text{cf}}^{(\ell)}(\cdot)$ is a learnable projection.
This top-down propagation injects global and regional routing context into fine-grid nodes, enabling local regions to reason about congestion beyond their immediate neighborhood.

\textbf{Gated integration.}
To balance local fidelity and global context, we adaptively combine the propagated hierarchical feature with the original fine-grid embedding using a learnable gate:
\begin{equation}
\vect{\gamma}_g
=
\sigmoid
\left(
\mat{W}_{\gamma}
\big[
\vect{h}_{g}^{(0)} \;\|\; \tilde{\vect{h}}_{g}^{(0)}
\big]
\right),
\label{eq:grid_gate}
\end{equation}
\begin{equation}
\vect{h}_{g}^{\text{out}}
=
\vect{\gamma}_g \odot \tilde{\vect{h}}_{g}^{(0)}
+
(1-\vect{\gamma}_g) \odot \vect{h}_{g}^{(0)},
\label{eq:grid_gate_fuse}
\end{equation}
where $\vect{h}_{g}^{(0)}$ is the original fine-grid embedding and $\odot$ denotes element-wise multiplication.
The gate $\vect{\gamma}_g$ controls the influence of multi-scale spatial context on each grid node, ensuring that the hierarchical representation enhances congestion reasoning without overwhelming fine-grained local information.

\subsection{Congestion Prediction and Training Objective}
\label{sec:training}

\begin{figure}[t]
    \centering
    \includegraphics[width=\linewidth]{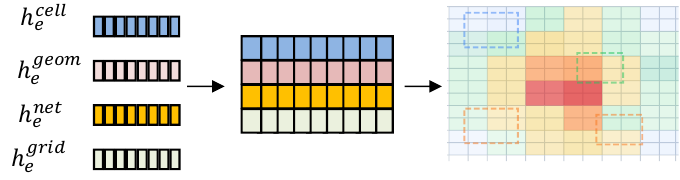}
    \caption{Embedding and readout. Embeddings from different node types (cell, geom, net, and grid) are concatenated and mapped through a readout layer to produce the final congestion heatmap.}
    \label{fig:embedding_readout}
\end{figure}

Figure~\ref{fig:embedding_readout} illustrates the final prediction stage after message passing. Separate prediction heads are applied to the final cell and grid embeddings to produce instance-level and region-level congestion estimates, enabling supervision at both hotspot and regional routing-pressure granularities.

\textbf{Weighted regression loss.}
Routing congestion is highly imbalanced: most cells and grid regions have near-zero congestion, while a small fraction forms critical hotspots. Standard mean squared error can therefore be dominated by low-congestion samples and lead to low-variance predictions. To emphasize congested regions, we adopt a weighted mean squared error:
\begin{equation}
    \mathcal{L}_{\text{wmse}} =
\frac{1}{|\mathcal{V}|}
\sum_{v \in \mathcal{V}}
w_v \, (\hat{c}_v - c_v)^2,
\end{equation}
where $c_v$ and $\hat{c}_v$ denote the ground-truth and predicted congestion values, and $w_v$ is a congestion-dependent weight. This weighting encourages the model to better distinguish high-congestion hotspots that are most relevant to downstream placement and routing decisions.

\textbf{Variance regularization and model selection.}
To further prevent prediction collapse, we add a variance regularization term that discourages near-constant outputs and preserves relative differences among congestion levels. The final objective combines cell-level and grid-level weighted regression losses with this regularization. Since raw loss values may still be dominated by low-congestion regions, model selection uses validation Spearman rank correlation, which better reflects hotspot ordering quality. Early stopping under this criterion prioritizes congestion ranking over average numerical error.

% \textbf{Variance regularization.}
% Although weighted regression mitigates label imbalance, it does not fully prevent prediction collapse. We therefore include a variance regularization term that penalizes insufficient output dispersion, discouraging near-constant predictions and preserving relative differences among congestion levels.

% \textbf{Optimization and early stopping.}
% The final objective combines cell-level and grid-level weighted regression losses with variance regularization. Since raw loss values can be dominated by low-congestion regions, model selection is based on validation Spearman rank correlation, which better reflects congestion ordering quality. Early stopping under this criterion prioritizes correct hotspot ranking over average numerical error.

\section{Experiment}
We conduct extensive experiments to evaluate VeriHGN on standard VLSI congestion prediction benchmarks. Our evaluation aims to answer four research questions.

\begin{itemize}[leftmargin=*]
    \item \textbf{RQ1:} How does VeriHGN compare with state-of-the-art congestion predictors?
    \item \textbf{RQ2:} What is the contribution of each proposed component?
    \item \textbf{RQ3:} How sensitive is the model to key hyperparameters?
    \item \textbf{RQ4:} How does VeriHGN generalize across design nodes?
\end{itemize}

\subsection{Experimental Setup}
\label{sec:exp_setup}

\textbf{Datasets.} We evaluate on CircuitNet~\cite{2024circuitnet}, an open-source dataset for machine learning applications in EDA, including three public datasets, ISPD-2015, CircuitNet-N14, and CircuitNet-N28.
Table~\ref{tab:dataset} presents an overview of the datasets used in our experiments. The benchmarks span both academic and industrial settings, with circuit sizes ranging from 35K to 1.5M cells and 38K to 1.6M nets, and include multiple placement and routing snapshots for robust evaluation.

\begin{table}[t]
\centering
\caption{Summary of datasets used in our experiments.}
\label{tab:dataset}
\small
\setlength{\tabcolsep}{3pt}
\begin{tabular}{lcccccc}
\toprule
Dataset & Tech. & \#Designs & \#Cells & \#Nets & \#Samples \\
\midrule
ISPD 2015 & -- & 20 & 100K--1.2M & 100K--1.3M & 533 \\
CircuitNet-N28 & 28nm & 10 & 35K--1.5M & 38K--1.6M & $\sim$3K \\
CircuitNet-N14 & 14nm & 8 & 35K--1.5M & 38K--1.6M & $\sim$2K \\
\bottomrule
\end{tabular}
\end{table}

\textbf{Data preprocessing.} For each dataset, we parse placement DEF files to extract cell coordinates, dimensions, and netlist topology. Ground-truth congestion maps, which are provided at design-specific global-routing resolutions, are area-averaged to the target grid resolution using bilinear downsampling so that full-chip congestion information is preserved. Congestion targets are normalized as $\log(1+x) / c_{\max}$, where $c_{\max}$ is the per-dataset maximum of $\log(1+x)$, reducing the effect of heavy-tailed congestion values while keeping targets approximately within $[0,1]$.

\textbf{Evaluation metrics.}
We evaluate congestion prediction using both rank-based correlation metrics and error-based metrics. Correlation metrics measure whether the model preserves congestion trends and hotspot ordering, while error metrics quantify the deviation between predicted and ground-truth congestion magnitudes.
\begin{table*}[t]
\centering
\small
\setlength{\tabcolsep}{3.2pt}
\renewcommand{\arraystretch}{1.05}
\caption{Overall comparison of cell-level and grid-level congestion prediction performance.}
\label{tab:main_results}
\begin{tabular}{llrrrrrrrrrr}
\toprule
\multirow{2}{*}{\textbf{Dataset}} 
& \multirow{2}{*}{\textbf{Model}}
& \multicolumn{5}{c}{\textbf{Cell-level}}
& \multicolumn{5}{c}{\textbf{Grid-level}} \\
\cmidrule(lr){3-7} \cmidrule(lr){8-12}
& & \textbf{MAE}$\downarrow$ 
& \textbf{RMSE}$\downarrow$ 
& \textbf{Pearson}$\uparrow$ 
& \textbf{Spearman}$\uparrow$ 
& \textbf{Kendall}$\uparrow$
& \textbf{MAE}$\downarrow$ 
& \textbf{RMSE}$\downarrow$ 
& \textbf{Pearson}$\uparrow$ 
& \textbf{Spearman}$\uparrow$ 
& \textbf{Kendall}$\uparrow$ \\
\midrule

\multirow{7}{*}{ISPD2015}
& GCN        & 0.041 & 0.047 & 0.347 & 0.398 & 0.321 & 0.038 & 0.049 & 0.299 & 0.273 & 0.163 \\
& GAT        & 0.039 & 0.045 & 0.361 & 0.385 & 0.337 & 0.036 & 0.046 & 0.304 & 0.273 & 0.179 \\
& CircuitGNN & 0.034 & 0.040 & 0.598 & 0.322 & 0.217 & 0.042 & 0.049 & 0.364 & 0.162 & 0.134 \\
& MIHC       & 0.029 & 0.034 & \textbf{0.687} & 0.689 & 0.447 & 0.032 & 0.039 & 0.503 & 0.271 & \textbf{0.227} \\
& ST-FPN     & 0.039 & 0.043 & 0.533 & 0.332 & 0.290 & 0.037 & 0.041 & 0.483 & 0.375 & 0.208 \\
& Lay-Net    & 0.034 & 0.037 & 0.625 & 0.400 & 0.324 & 0.041 & 0.044 & 0.671 & \textbf{0.411} & 0.198 \\
& VeriHGN    & \textbf{0.025} & \textbf{0.031} & 0.663 & \textbf{0.692} & \textbf{0.465}
              & \textbf{0.029} & \textbf{0.035} & \textbf{0.677} & 0.308 & 0.211 \\
\midrule

\multirow{7}{*}{CircuitNet-N14}
& GCN        & 0.051 & 0.058 & 0.341 & 0.336 & 0.309 & 0.048 & 0.060 & 0.281 & 0.248 & 0.226 \\
& GAT        & 0.050 & 0.057 & 0.351 & 0.342 & 0.298 & 0.047 & 0.059 & 0.289 & 0.255 & 0.219 \\
& CircuitGNN & 0.046 & 0.053 & 0.402 & 0.378 & 0.316 & 0.054 & 0.062 & 0.283 & 0.343 & 0.318 \\
& MIHC       & 0.036 & 0.041 & 0.524 & 0.512 & 0.447 & 0.034 & 0.043 & 0.417 & \textbf{0.358} & 0.294 \\
& ST-FPN     & 0.041 & 0.049 & 0.411 & 0.308 & 0.299 & 0.037 & 0.045 & 0.410 & 0.320 & 0.312 \\
& Lay-Net    & \textbf{0.031} & \textbf{0.033} & \textbf{0.566} & 0.344 & 0.301
              & \textbf{0.033} & \textbf{0.037} & 0.484 & 0.348 & \textbf{0.329} \\
& VeriHGN    & 0.038 & 0.043 & 0.543 & \textbf{0.513} & \textbf{0.461}
              & 0.035 & 0.044 & \textbf{0.489} & 0.331 & 0.322 \\
\midrule

\multirow{7}{*}{CircuitNet-N28}
& GCN        & 0.045 & 0.051 & 0.516 & 0.438 & 0.319 & 0.043 & 0.053 & 0.408 & 0.308 & 0.231 \\
& GAT        & 0.046 & 0.051 & 0.513 & 0.447 & 0.316 & 0.044 & 0.055 & 0.414 & 0.316 & 0.234 \\
& CircuitGNN & 0.039 & 0.040 & 0.609 & 0.500 & 0.368 & 0.047 & 0.048 & 0.387 & 0.236 & 0.183 \\
& MIHC       & \textbf{0.036} & \textbf{0.039} & 0.661 & 0.522 & 0.396 & \textbf{0.034} & \textbf{0.041} & 0.551 & 0.374 & 0.286 \\
& ST-FPN     & 0.039 & 0.041 & 0.588 & 0.519 & 0.384 & 0.037 & 0.043 & 0.536 & 0.371 & 0.277 \\
& Lay-Net    & 0.037 & 0.041 & 0.617 & 0.522 & 0.399 & 0.035 & \textbf{0.041} & 0.558 & 0.379 & 0.285 \\
& VeriHGN    & 0.037 & 0.040 & \textbf{0.683} & \textbf{0.548} & \textbf{0.414}
              & \textbf{0.034} & \textbf{0.041} & \textbf{0.568} & \textbf{0.401} & \textbf{0.308} \\
\bottomrule
\end{tabular}
\end{table*}
\textit{Pearson correlation coefficient ($r$)} measures the linear relationship between predictions $\hat{\vect{y}} = (\hat{y}_1, \ldots, \hat{y}_n)$ and targets $\vect{y} = (y_1, \ldots, y_n)$:
\begin{equation}
r = \frac{\sum_{i=1}^{n} (\hat{y}_i - \bar{\hat{y}})(y_i - \bar{y})}{\sqrt{\sum_{i=1}^{n} (\hat{y}_i - \bar{\hat{y}})^2} \cdot \sqrt{\sum_{i=1}^{n} (y_i - \bar{y})^2}}
\label{eq:pearson}
\end{equation}
where $\bar{\hat{y}}$ and $\bar{y}$ are the respective means. Pearson $r \in [-1,1]$ captures whether predictions scale linearly with true congestion values, but it is sensitive to outliers and magnitude distortion.

\textit{Spearman rank correlation ($\rho_s$)} evaluates monotonic agreement by computing Pearson correlation over ranked predictions and targets:
\begin{equation}
\rho_s = r\big(\text{rank}(\hat{\vect{y}}),\; \text{rank}(\vect{y})\big).
\label{eq:spearman}
\end{equation}
Unlike Pearson correlation, Spearman $\rho_s$ focuses on whether the model correctly orders cells or grid bins from low to high congestion. This is important for layout verification, where engineers often need to prioritize the most critical hotspots rather than precisely match every congestion magnitude.

\textit{Kendall rank correlation ($\tau$)} measures pairwise ordering consistency:
\begin{equation}
\tau = \frac{n_c - n_d}{\binom{n}{2}},
\label{eq:kendall}
\end{equation}
where $n_c$ and $n_d$ are the numbers of concordant and discordant pairs, respectively. Kendall $\tau \in [-1,1]$ provides a conservative ranking metric and is useful when many cells or grid bins have tied or near-zero congestion values.

\textit{Mean absolute error (MAE)} quantifies the average absolute difference between predicted and ground-truth congestion:
\begin{equation}
\mathrm{MAE} = \frac{1}{n} \sum_{i=1}^{n} \lvert \hat{y}_i - y_i \rvert .
\label{eq:mae}
\end{equation}
MAE reflects the typical magnitude of prediction error and is less sensitive to extreme outliers than squared-error metrics.

\textit{Root mean squared error (RMSE)} emphasizes larger deviations by squaring errors before averaging:
\begin{equation}
\mathrm{RMSE} = \sqrt{\frac{1}{n} \sum_{i=1}^{n} (\hat{y}_i - y_i)^2}.
\label{eq:rmse}
\end{equation}
RMSE penalizes large congestion mispredictions more heavily and complements MAE by highlighting severe local errors. Together, these metrics provide a comprehensive evaluation of congestion prediction performance: Pearson measures value-level linear accuracy, Spearman and Kendall focus on ranking quality and hotspot ordering, while MAE and RMSE quantify absolute prediction errors. 
All metrics are computed separately for cell-level and grid-level predictions.

\textbf{Implementation details.}
VeriHGN is implemented in PyTorch 2.1~\cite{paszke2019pytorch} with PyTorch Geometric 2.4~\cite{fey2019fast}. Experiments are conducted on a single NVIDIA L40S GPU with CUDA 11.8. We train using AdamW with learning rate $5 \times 10^{-4}$ and weight decay $10^{-4}$, together with a StepLR scheduler with step size 50 and $\gamma = 0.5$, for up to 200 epochs. Early stopping with patience 20 is based on the average validation Spearman correlation $\bar{\rho}_s = (\rho_s^{\text{cell}} + \rho_s^{\text{grid}}) / 2$. We use Spearman rather than validation loss because low MSE can be misleading under sparse congestion distributions, where near-constant predictions may achieve low loss but poor ranking quality.

\subsection{Congestion Prediction Comparison}

We compare VeriHGN with six baselines: two general-purpose GNNs, GCN~\cite{kipf2016semi} and GAT~\cite{velivckovic2017graph}; two EDA-specific graph-based predictors, CircuitGNN~\cite{yang2022versatile} and MIHC~\cite{zhangmihc}; and two geometry-based methods, ST-FPN~\cite{cai2025st} and Lay-Net~\cite{zou2025lay}. Lay-Net is evaluated using its open-source implementation, while ST-FPN is reimplemented following the original paper. This comparison covers both topology-aware graph models and dense layout-based models, allowing us to evaluate the trade-off between structural reasoning and spatial regression.

Table~\ref{tab:main_results} reports cell-level and grid-level results on ISPD2015, CircuitNet-N14, and CircuitNet-N28. Overall, VeriHGN achieves best or near-best performance, especially on rank-based metrics. At the cell level, VeriHGN obtains the best Spearman and Kendall correlations on all three datasets, indicating a stronger ability to preserve congestion severity ordering. This is important for practical layout verification, where prioritizing critical hotspots is often more useful than minimizing average prediction error alone. Compared with GCN, GAT, and CircuitGNN, the gains demonstrate the benefit of explicitly modeling heterogeneous cell-net-grid interactions. Compared with MIHC, VeriHGN achieves stronger ranking quality in most settings, suggesting that unified heterogeneous message passing is more effective than loosely coupled multi-view fusion for hotspot ordering.

The comparison with geometry-based baselines reveals a clear trade-off. Lay-Net and ST-FPN remain highly competitive on direct value-regression metrics, especially on CircuitNet-N14, where Lay-Net achieves the best cell-level MAE, RMSE, and Pearson correlation. This confirms the strength of dense spatial encoders for layout-level congestion magnitude prediction. In contrast, VeriHGN is more competitive on ranking metrics and achieves the strongest overall performance on CircuitNet-N28, including the best Pearson, Spearman, and Kendall correlations at both cell and grid levels. These results suggest that geometry-based models excel at dense spatial regression, while VeriHGN better captures structural routing demand and hotspot ordering through explicit heterogeneous graph modeling.

At the grid level, the results are more nuanced. VeriHGN achieves the best grid-level MAE, RMSE, and Pearson on ISPD2015 and the strongest grid-level correlation metrics on CircuitNet-N28. On CircuitNet-N14, Lay-Net and MIHC remain slightly stronger on several grid-level metrics, indicating that grid-level prediction is more sensitive to spatial resolution and layout-specific distributions. Nevertheless, VeriHGN maintains competitive grid-level accuracy while consistently improving cell-level hotspot ranking, demonstrating a robust balance between fine-grained structural reasoning and regional congestion prediction.

\subsection{Ablation Study}
\label{sec:ablation}

To quantify the contribution of each proposed component, we conduct a systematic ablation study on CircuitNet-N28, which is a representative benchmark with varying scale designs. In our evaluation, starting from the full VeriHGN model, we remove or replace one component at a time and report cell-level and grid-level performance. The results are summarized in Table~\ref{tab:ablation}.

Several key observations emerge from the ablation results. First, hierarchical grid representation is critical for grid-level prediction.
Removing the multi-resolution grid hierarchy causes the largest drop in grid-level Spearman correlation ($0.401 \to 0.342$, $-14.7\%$), confirming that multi-scale spatial aggregation is essential for capturing regional routing pressure that spans beyond fine-grid neighborhoods. Second, disabling grid-to-net messages leads to substantial degradation in both cell-level Spearman ($0.548 \to 0.503$) and grid-level Spearman ($0.401 \to 0.356$). This result validates our design choice of propagating localized routing-resource information from grid tiles to net representations, which enables nets to be aware of the spatial congestion context along their routing paths. Furthermore, replacing the hand-crafted features with minimal coordinate and connectivity features yields the largest overall performance drop, with cell-level Spearman falling from $0.548$ to $0.478$. This highlights that explicitly encoding routing-demand signals such as HPWL statistics, pin density, and type-related attributes is far more sample-efficient than relying on deep message passing to implicitly recover these signals, particularly given the limited number of training designs in current EDA benchmarks.

\subsection{Hyperparameter Sensitivity}
\label{sec:sensitivity}

\begin{figure}[t]
    \centering
    \includegraphics[width=\linewidth]{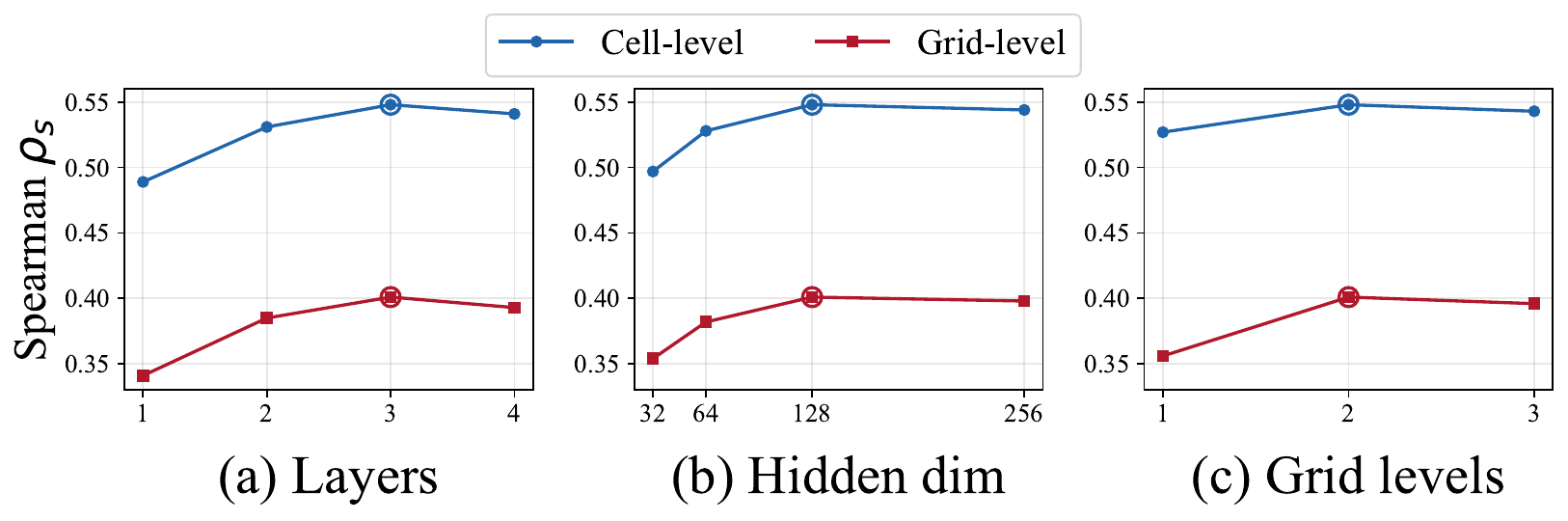}
    \caption{Sensitivity of VeriHGN to key hyperparameters on CircuitNet-N28: (a) number of message-passing layers $L$, (b) hidden dimension $d$, and (c) number of hierarchical grid levels $K$. Circles highlight the optimal configuration ($L=3$, $d=128$, $K=2$).}
    \label{fig:sensitivity}
\end{figure}

We analyze the sensitivity of VeriHGN to three key hyperparameters: the number of message-passing layers $L$, hidden dimension $d$, and the number of hierarchical grid levels $K$. Experiments are conducted on CircuitNet-N28 with all other settings fixed. Figure~\ref{fig:sensitivity} summarizes the performance trends under different configurations. Overall, VeriHGN exhibits stable behavior across a wide range of hyperparameters, indicating that the proposed architecture does not rely on aggressive tuning to achieve strong performance.

\begin{table}[t]
\caption{Ablation study on CircuitNet-N28. "None" denotes the complete VeriHGN model. Each row removes or replaces one component.}
\label{tab:ablation}
\centering
\small
\setlength{\tabcolsep}{3pt}
\begin{tabular}{l|ccc|ccc}
\toprule
 & \multicolumn{3}{c|}{Cell-level} & \multicolumn{3}{c}{Grid-level} \\
Remove Variant & MAE$\downarrow$ & Spear.$\uparrow$ & Kend.$\uparrow$ & MAE$\downarrow$ & Spear.$\uparrow$ & Kend.$\uparrow$ \\
\midrule
None           & \textbf{0.037} & \textbf{0.548} & \textbf{0.414} & \textbf{0.034} & \textbf{0.401} & \textbf{0.308} \\
\midrule
Hierarchical Grid  & 0.039 & 0.511 & 0.383 & 0.038 & 0.342 & 0.261 \\
 Grid-Net MP        & 0.040 & 0.503 & 0.371 & 0.037 & 0.356 & 0.274 \\
 Geom. Cell--Cell MP & 0.038 & 0.527 & 0.396 & 0.035 & 0.381 & 0.290 \\
 Gated Aggregation  & 0.038 & 0.531 & 0.399 & 0.035 & 0.387 & 0.295 \\
 Enriched Features  & 0.042 & 0.478 & 0.354 & 0.039 & 0.331 & 0.252 \\
 Weighted Loss      & 0.041 & 0.462 & 0.341 & 0.036 & 0.348 & 0.267 \\
 Variance Reg.      & 0.039 & 0.491 & 0.368 & 0.036 & 0.363 & 0.278 \\
\bottomrule
\end{tabular}
\end{table}

% \begin{table}[t]
% \caption{Sensitivity to the number of message-passing layers $L$, hidden dimension $d$, and hierarchical grid levels $K$ on CircuitNet-N28.}
% \label{tab:sensitivity}
% \centering
% \small
% \setlength{\tabcolsep}{3pt}
% \begin{tabular}{cc|ccc|ccc}
% \toprule
%  & & \multicolumn{3}{c|}{Cell-level} & \multicolumn{3}{c}{Grid-level} \\
% \multicolumn{2}{c|}{Hyperparameter} & MAE$\downarrow$ & Spear.$\uparrow$ & Kend.$\uparrow$ & MAE$\downarrow$ & Spear.$\uparrow$ & Kend.$\uparrow$ \\
% \midrule
% \multirow{4}{*}{Layers $L$}
%  & 1 & 0.043 & 0.489 & 0.358 & 0.039 & 0.341 & 0.254 \\
%  & 2 & 0.039 & 0.531 & 0.397 & 0.036 & 0.385 & 0.291 \\
%  & \textbf{3} & \textbf{0.037} & \textbf{0.548} & \textbf{0.414} & \textbf{0.034} & \textbf{0.401} & \textbf{0.308} \\
%  & 4 & 0.038 & 0.541 & 0.406 & 0.035 & 0.393 & 0.299 \\
% \midrule
% \multirow{4}{*}{Hidden dim $d$}
%  & 32  & 0.044 & 0.497 & 0.367 & 0.040 & 0.354 & 0.264 \\
%  & 64  & 0.040 & 0.528 & 0.395 & 0.037 & 0.382 & 0.289 \\
%  & \textbf{128} & \textbf{0.037} & \textbf{0.548} & \textbf{0.414} & \textbf{0.034} & \textbf{0.401} & \textbf{0.308} \\
%  & 256 & 0.037 & 0.544 & 0.410 & 0.034 & 0.398 & 0.304 \\
% \midrule
% \multirow{3}{*}{Grid levels $K$}
%  & 1 & 0.039 & 0.527 & 0.393 & 0.037 & 0.356 & 0.268 \\
%  & \textbf{2} & \textbf{0.037} & \textbf{0.548} & \textbf{0.414} & \textbf{0.034} & \textbf{0.401} & \textbf{0.308} \\
%  & 3 & 0.037 & 0.543 & 0.409 & 0.035 & 0.396 & 0.301 \\
% \bottomrule
% \end{tabular}
% \end{table}

\textbf{Message-passing depth.}
Performance improves steadily from $L=1$ to $L=3$, with both MAE and ranking metrics reaching their optimum. At $L=4$, a slight degradation is observed, consistent with the well-known oversmoothing~\cite{chen2020measuring} phenomenon in graph neural networks where excessive propagation causes node embeddings to converge. The relatively shallow optimal depth ($L=3$) also reflects the effectiveness of our enriched features, which reduce the model's reliance on deep propagation to recover routing-demand signals.

\textbf{Hidden dimension.}
Increasing $d$ from 32 to 128 yields consistent gains across all metrics. Beyond $d=128$, performance saturates, suggesting that 128 dimensions provide sufficient representational capacity for the heterogeneous node and edge types. The slight degradation at $d=256$ may indicate mild overfitting given the limited number of training designs.

\textbf{Number of grid levels.}
A single grid level ($K=1$) captures only local spatial patterns, resulting in notably weaker grid-level Spearman correlation ($0.356$). Adding one coarser level ($K=2$) yields substantial improvement by enabling regional routing context to inform fine-grid predictions. A third level ($K=3$) provides no further benefit, likely because the coarsest grid becomes too spatially aggregated for the circuit sizes in CircuitNet-N28.

\subsection{Cross-design Generalization and Few-shot Adaptation}
\label{sec:cross_tech}
An important practical goal for learning-based congestion prediction is cross-design generalization, since collecting labeled routing data for every new process is expensive. We first evaluate
zero-shot cross-design transfer by training on one design node and
directly testing on another without fine-tuning. We compare against
MIHC under the same transfer protocol.

Table~\ref{tab:cross_tech} reports zero-shot cross-design transfer between CircuitNet-N14 and CircuitNet-N28. 
Both MIHC and VeriHGN suffer clear degradation under transfer, reflecting the distribution shift in cell libraries, routing resources, and congestion patterns across technology nodes. 
Nevertheless, VeriHGN consistently outperforms MIHC in both transfer directions.

In the N28$\rightarrow$N14 direction, VeriHGN retains 43.7\% of its in-domain Spearman correlation, compared with 32.8\% for MIHC. 
In the reverse direction, VeriHGN achieves 47.1\% retention versus 38.5\% for MIHC. 
Although zero-shot transfer is not yet sufficient for practical deployment, these results suggest that VeriHGN learns more transferable routing-demand representations and provides a stronger initialization for target-domain adaptation.
\begin{table}[t]
\caption{Cross-design generalization (cell-level). $\Delta\rho_s$ denotes the relative Spearman retention under transfer: $\rho_s^{\text{transfer}} / \rho_s^{\text{in-domain}}$.}
\label{tab:cross_tech}
\centering
\small
\setlength{\tabcolsep}{4pt}
\begin{tabular}{ll|cccc|c}
\toprule
Train $\to$ Test & Model & MAE$\downarrow$ & Pears.$\uparrow$ & Spear.$\uparrow$ & Kend.$\uparrow$ & $\Delta\rho_s$ \\
\midrule
N14 $\to$ N14 & MIHC    & 0.036 & 0.524 & 0.512 & 0.447 & -- \\
N14 $\to$ N14 & VeriHGN & 0.038 & 0.543 & 0.513 & 0.461 & -- \\
N28 $\to$ N28 & MIHC    & 0.036 & 0.661 & 0.522 & 0.396 & -- \\
N28 $\to$ N28 & VeriHGN & 0.037 & 0.663 & 0.548 & 0.414 & -- \\
\midrule
N28 $\to$ N14 & MIHC    & 0.062 & 0.198 & 0.168 & 0.114 & 32.8\% \\
N28 $\to$ N14 & VeriHGN & 0.056 & 0.261 & 0.224 & 0.159 & 43.7\% \\
N14 $\to$ N28 & MIHC    & 0.058 & 0.278 & 0.201 & 0.142 & 38.5\% \\
N14 $\to$ N28 & VeriHGN & 0.052 & 0.347 & 0.258 & 0.187 & 47.1\% \\
\bottomrule
\end{tabular}
\end{table}

These results also show that cross-design generalization remains an open challenge for learning-based congestion prediction. Although VeriHGN's unified representation improves transfer robustness, practical deployment on a new design node is likely to require limited in-domain supervision. 

To examine whether limited target-domain labels can close this gap, we further conduct a few-shot fine-tuning analysis. We pretrain VeriHGN on the source node, initialize the target-node model from the zero-shot checkpoint, and fine-tune it using increasing fractions of labeled target samples. The fine-tuned model is evaluated on the held-out target test set, with cell-level Spearman reported in Table~\ref{tab:fewshot_transfer}. 

Few-shot fine-tuning substantially improves transfer performance. For N28$\rightarrow$N14, Spearman increases from 0.220 under zero-shot transfer to 0.330 with 10\% target data and 0.490 with 50\% target data, recovering 96.1\% of fully supervised non-transfer performance. A similar trend appears for N14$\rightarrow$N28, where 50\% target data improves Spearman from 0.258 to 0.516. These results suggest that transfer difficulty is mainly caused by domain shift rather than a failure of the learned representation. Performance saturates after 50--70\% target data, indicating diminishing returns from additional labels; thus, modest target-domain supervision is sufficient to close most of the gap to fully supervised training.

\subsection{Runtime and Memory Analysis}
\label{sec:runtime_analysis}

To further understand the practical cost of heterogeneous graph-based congestion prediction, we report the runtime and peak GPU memory of VeriHGN, MIHC, and CircuitGNN in Table~\ref{tab:runtime_memory}. We measure graph construction time on representative small, medium, and large designs, together with total training time, per-epoch training time, inference latency, and peak GPU memory. The scalability challenge is consistent with prior observations that GNN workloads are often limited by irregular memory access, sparse aggregation, and graph-dependent execution patterns~\cite{ma2019neugraph,wang2021gnnadvisor,huang2021understanding,wu2021seastar}.
\begin{table}[t]
\centering
\scriptsize
\setlength{\tabcolsep}{3.2pt}
\renewcommand{\arraystretch}{1.08}
\caption{Few-shot fine-tuning sensitivity under cross-design transfer. 
The reported metric is cell-level Spearman correlation.}
\label{tab:fewshot_transfer}
\begin{tabular*}{\columnwidth}{@{\extracolsep{\fill}}lccccccc@{}}
\toprule
\textbf{Transfer} 
& \textbf{0\%} 
& \textbf{10\%} 
& \textbf{30\%} 
& \textbf{50\%} 
& \textbf{70\%} 
& \textbf{90\%} 
& \textbf{Non-trans.} \\
\midrule
N28 $\rightarrow$ N14 & 0.224 & 0.330 & 0.400 & \textbf{0.490} & 0.500 & 0.510 & 0.513 \\
N14 $\rightarrow$ N28 & 0.258 & 0.353 & 0.421 & \textbf{0.516} & 0.538 & 0.544 & 0.548 \\
\bottomrule
\end{tabular*}
\end{table}
The results show that VeriHGN incurs higher preprocessing and memory overhead than CircuitGNN because it constructs richer heterogeneous graphs with multiple node and relation types. Its longer training time on CircuitNet-N14 and CircuitNet-N28 reflects the additional cost of relation-specific message passing and structural fusion. Thus, the improved modeling capacity of VeriHGN comes with a clear computational cost. Compared with MIHC, however, VeriHGN exhibits a comparable efficiency profile. On ISPD2015, it achieves nearly the same total training time while reducing per-epoch time from 29.1s to 22.2s and inference latency from 14.6s to 9.7s. On CircuitNet-N14 and CircuitNet-N28, VeriHGN has similar total training time to MIHC, with only slightly higher per-epoch cost. Its inference latency is higher on CircuitNet-N14 but lower on CircuitNet-N28, suggesting that runtime differences depend on graph scale and dataset structure rather than being uniformly worse. VeriHGN generally uses more peak memory, which is expected because it maintains heterogeneous node, edge, and relation-specific representations during training. Overall, these results demonstrate an accuracy--efficiency trade-off. Geometry-based methods such as Lay-Net and ST-FPN can benefit from dense tensor operations and avoid graph construction overhead, while graph-based methods require additional preprocessing and sparse message passing. VeriHGN is therefore slower than lightweight or geometry-based baselines during training, but its inference latency remains practical across all datasets. This makes VeriHGN suitable when hotspot ranking, heterogeneous structural reasoning, and cross-domain circuit interactions are important, whereas dense layout-based models may remain preferable when runtime or direct value-regression error is the primary concern.

\section{Discussion}
\begin{table}[t]
\centering
\scriptsize
\setlength{\tabcolsep}{2.2pt}
\renewcommand{\arraystretch}{1.08}
\caption{Runtime and memory comparison across datasets.}
\label{tab:runtime_memory}
\begin{tabular*}{\columnwidth}{@{\extracolsep{\fill}}llccccc@{}}
\toprule
\textbf{Model} 
& \textbf{Dataset}
& \textbf{GC (S,M,L) (s)}
& \textbf{Train (min)}
& \textbf{Epoch (s)}
& \textbf{Inf. (s)}
& \textbf{Mem. (GB)} \\
\midrule

\multirow{3}{*}{CircuitGNN}
& ISPD2015 & (8.0, 180.3, 286.4)   & 41.3  & 17.0  & 9.0  & 14.0 \\
& N14      & (2.5, 160.7, 303.3)   & 180.3 & 18.0  & 13.3 & 21.0 \\
& N28      & (2.7, 177.9, 300.0)   & 161.7 & 15.0  & 11.3 & 17.0 \\
\midrule

\multirow{3}{*}{MIHC}
& ISPD2015 & (2.8, 215.4, 678.6)   & 97.6  & 29.1  & 14.6 & 18.45 \\
& N14      & (3.9, 283.5, 684.0)   & 928.6 & 300.2 & 7.3  & 28.56 \\
& N28      & (3.7, 271.5, 584.0)   & 842.9 & 283.6 & 26.0 & 38.56 \\
\midrule

\multirow{3}{*}{VeriHGN}
& ISPD2015 & (2.9, 291.5, 743.2)   & 97.3  & 22.2  & 9.7  & 30.2 \\
& N14      & (3.3, 248.3, 747.0)   & 917.7 & 312.2 & 27.0 & 37.2 \\
& N28      & (2.7, 218.5, 647.0)   & 846.0 & 287.3 & 19.4 & 38.7 \\
\bottomrule
\end{tabular*}

\vspace{0.25em}
\parbox{\columnwidth}{\scriptsize
\textit{Note:} GC denotes graph construction time. 
S, M, and L denote small, medium, and large designs, respectively. 
N14 and N28 denote CircuitNet-N14 and CircuitNet-N28.
}
\vspace{-0.5em}
\end{table}

VeriHGN complements topology- and geometry-centric congestion predictors by incorporating heterogeneous graph structures on top of layout-derived features. By jointly modeling cells, nets, and hierarchical grid regions, it captures structural routing demand and multi-scale spatial constraints that are difficult to represent with rasterized layout features alone. This enables strong rank-based performance, especially at the cell level, where accurate hotspot prioritization is critical for verification.

Our results also reveal important trade-offs. Geometry-based methods such as Lay-Net and ST-FPN remain competitive on direct error metrics, while VeriHGN provides stronger structural reasoning and hotspot ordering. Although zero-shot cross-design transfer remains challenging, few-shot fine-tuning substantially improves performance, suggesting useful transferable representations. While graph construction and relation-specific message passing increase training and memory cost, VeriHGN maintains practical inference latency, making this trade-off reasonable for deployment-oriented congestion verification.

\section{Conclusion}
\label{sec:conclusion}

We presented VeriHGN, an early-stage congestion prediction framework that models circuit components and spatial grids as a unified heterogeneous graph, capturing interactions between netlist topology and physical layout geometry. Through enriched features, relation-specific message passing, hierarchical grid learning, and congestion-weighted training, VeriHGN achieves competitive or superior performance on ISPD2015, CircuitNet-N14, and CircuitNet-N28. Future work will integrate geometry-based encoders, extend to timing-aware congestion prediction, and improve cross-technology transfer via domain adaptation.
\section*{Acknowledgment }
\noindent This work was supported in part by National Science Foundation grants 2331301, 2508118, 2516003, 2419843.
The views, opinions, and/or findings expressed in this material are those of the authors and should not be interpreted as representing the official views of the National Science Foundation, or the U.S. Government.
\clearpage

\balance
\bibliographystyle{ACM-Reference-Format}
\bibliography{refs}

\end{document}